\documentclass[fleqn,usenatbib]{mnras}
\usepackage{newtxtext,newtxmath}
\usepackage[T1]{fontenc}
\usepackage{graphicx}
\usepackage{bm}
\usepackage{amsmath}
\usepackage{natbib}
\usepackage{booktabs}
\usepackage{capt-of}
\usepackage{color}
\usepackage{xcolor}
\usepackage{orcidlink}
\usepackage{hyperref}
\usepackage[normalem]{ulem}

\newcommand{\Msun}{{M_{\odot}}}
\newcommand{\Mni}{{M_{\rm Ni}}}
\newcommand{\Mej}{{M_{\rm ej}}}
\newcommand{\Ek}{{E_{\rm k}}}
\newcommand{\Lbol}{{L_{\rm bol}}}
\newcommand{\program}[1]{\textsc{#1}}
\newcommand{\snmix}{\program{snmix}}
\newcommand{\citeg}[1]{\citep[e.g.,][]{#1}}
\newcommand{\be}{\begin{equation}}
\newcommand{\ee}{\end{equation}}

\defcitealias{SarinAlvey_inprep}{Sarin \& Alvey, in prep.}

\title[Physical Type Ia populations and standardisation]{A physical population model for Type Ia supernovae: implications for standardisation}

\author[N.~Sarin et al.]{Nikhil Sarin$^{1,2}$\thanks{E-mail: nsarin.astro@gmail.com}%
\orcidlink{0000-0003-2700-1030}, Matthew Grayling$^{1,2}$ and Lisa Kelsey$^{1,2}$%
\orcidlink{0000-0003-0313-0487} \\
$^{1}$Kavli Institute for Cosmology, University of Cambridge, Madingley Road, CB3 0HA, UK\\
$^{2}$Institute of Astronomy, University of Cambridge, Madingley Road, CB3 0HA, UK\\}

\date{Accepted XXX. Received YYY; in original form ZZZ}

\pubyear{\the\year{}}

\begin{document}
\label{firstpage}
\pagerange{\pageref{firstpage}--\pageref{lastpage}}
\maketitle

\begin{abstract}
The inferred ejecta and $^{56}$Ni masses of Type~Ia supernova (SN~Ia) from photometric analyses depend strongly on the assumed spectral energy distribution (SED). Here, we introduce a physical population framework separating explosion physics from spectral evolution. A multi-shell radiation-diffusion model, \snmix{}, computes each SN's bolometric light curve from its ejecta mass, $^{56}$Ni mass, and mixing extent, while a learned, energy-normalised SED redistributes that luminosity across wavelength without adding or removing energy. We train the SED on Carnegie Supernova Project optical and near-infrared photometry, and apply the model to 845 normal SNe~Ia at $z<0.06$ from the second Zwicky Transient Facility (ZTF) data release. Against multi-code radiative-transfer benchmarks, \snmix{} recovers $^{56}$Ni masses within $2$--$6\%$, inside the $11$--$27\%$ inter-code spread. For ZTF we find per-event medians of $\Mni = 0.57\,\Msun$ and $\Mej = 1.33\,\Msun$ (hierarchical population medians $0.57$ and $1.39\,\Msun$). SNe in more massive hosts have ejecta lighter by $0.20\pm0.03\,\Msun$, which disappears at fixed SALT3 stretch ($+0.04\pm0.03\,\Msun$). Standardising peak rest-frame $B$ magnitudes with three distance-free physical properties ($\Mej$, host $A_V$, and peak intrinsic $(B-V)_0$ colour) matches SALT3 scatter on 727 events ($0.177$ vs.\ $0.180$\,mag weighted rms; $0.161$ vs.\ $0.160$\,mag NMAD) and lowers the host-mass step from $-0.093\pm0.014$ to $-0.073\pm0.013$\,mag (paired reduction $0.020$\,mag); adding $(r-i)_0$ with diffusion time achieves $0.173$\,mag weighted rms ($0.152$\,mag NMAD) with a $-0.088\pm0.013$\,mag step.
\end{abstract}


\section{Introduction}
\label{sec:intro}

Type~Ia supernovae (SNe~Ia) are powered by the radioactive decay of $^{56}$Ni synthesised during the thermonuclear explosion of a white dwarf~\citep{Colgate1969}, and the rise, peak, and post-peak decline of their bolometric light curves encode how much $^{56}$Ni was produced and how much ejecta that radiation must diffuse through~\citep{Arnett1982}. Despite decades of study, the relative contribution of competing progenitor and explosion channels, such as near-Chandrasekhar-mass delayed detonations~\citep{Khokhlov1991} versus sub-Chandrasekhar-mass double detonations~\citep{Fink2010,Sim2010,Shen2018}, remains an open question~\citeg{Hillebrandt2013,Maoz2014,Ruiter2025}.
Bolometric analyses of nearby samples have used light-curve diffusion physics to infer ejecta and $^{56}$Ni masses for individual events, finding a wide mass distribution extending below the Chandrasekhar limit and arguing that $25$--$50\%$ of normal SNe~Ia are consistent with sub-Chandrasekhar-mass progenitors~\citep{Scalzo2014a,Scalzo2014b,Scalzo2019}. 
The same light curves underpin SN~Ia cosmology, where empirical standardisation by light-curve width and colour~\citep{Phillips1993,Riess1996,Tripp1998} reduces the luminosity dispersion to $\sim 0.15$\,mag but leaves a persistent residual dependence on the properties of the host galaxy, most famously the host stellar mass~\citeg{Kelly2010,Sullivan2010,Lampeitl2010,Childress2013,Rigault2020,Kelsey2021,Kelsey2023}.

However, the bolometric light curve is not directly observed. Inferring explosion parameters from broadband photometry requires a spectral energy distribution (SED) or bolometric correction to map the physical luminosity into the observed passbands, a misspecification in that mapping can directly lead to biased explosion parameters that may be hard to diagnose. For example, a light-curve fit can reproduce the photometry well while returning biased physical parameters, because a misspecified SED can trade intrinsic colour against host dust, distance, or $^{56}$Ni mass. 
\citet{Sarin2026b} performed the largest analysis fitting a simple one-zone radioactive model to a photometric sample of 2,205 ZTF SNe~Ia, using a wavelength- and time-dependent opacity with an otherwise blackbody SED implemented in \textsc{Redback}~\citep{Sarin2024}. The inferred explosion properties, especially the $^{56}$Ni mass, showed strong sensitivity to this SED assumption, while other quantities like ejecta mass and velocity were less affected. 
In particular, the $^{56}$Ni mass of the overall population shifted from a median of $\approx 1.2\,\Msun$ across the full photometric sample to a lower limit of $0.45\,\Msun$ when inferred via reconstructed pseudo-bolometric fits. Such a discrepancy can arise either from the bolometric model or from the simplified SED prescription, for example when an optical-only blackbody over-extrapolates flux into the unobserved ultraviolet below $3000$\,\AA, where iron-group line blanketing strongly suppresses real SN~Ia emission.

\citet{Sarin2026b} also found host-dependent $^{56}$Ni masses whose amplitude depended on the assumed spectral model. 
Empirical light-curve models such as SALT2 and SALT3~\citep{Guy2007,Kenworthy2021} and hierarchical optical-to-near-infrared SED models such as BayeSN~\citep{Mandel2011,Mandel2017,Mandel2022, Grayling2024} learn flexible spectral surfaces directly from data, but their latent parameters carry no explicit explosion physics. Combining this empirical SED with a bolometric model offers a route to alleviate the limitations of the physical modelling while providing a physically-motivated anchor to the agnosticism of data-driven approaches. Although this comes at the cost of interpretability between a simplified blackbody SED or a wavelength and temperature-dependent opacity modification as adopted in~\citet{Sarin2026b} compared to an empirical SED. 

The second ZTF SN~Ia data release~\citep{Rigault2025a} supplies well-sampled multi-band light curves for 845 spectroscopically normal SNe~Ia in the nearly volume-complete $z<0.06$ regime, alongside host-galaxy stellar masses, colours, and Si\,\textsc{ii} expansion velocities for nearly half the sample~\citep{Burgaz2025a}. Recent advances in GPU-accelerated transient inference (\citealp{Grayling2026}; \citetalias{SarinAlvey_inprep}) now make jointly inferring explosion physics and spectral evolution practical at scale, providing direct routes to improving both the SED prescription and the bolometric model.
The Vera C. Rubin Observatory~\citep{Ivezic2019} and the Nancy Grace Roman Space Telescope~\citep{Hounsell2018,Rose2021} will expand current samples by orders of magnitude, making scalable, physical population inference essential. 

We develop a framework that combines a physical bolometric engine with a learned spectral model while enforcing a strict energy-conservation boundary between the two. A differentiable multi-shell radiation-diffusion model, \snmix{}~\citep{Sarin2026a}, determines each SN's bolometric luminosity from its explosion parameters, while a learned tensor-spline SED centred on the \citet{Hsiao2007} template is normalised to unit integral at every phase provides the description of that power across wavelength. 
The SED can therefore redistribute the available luminosity across wavelength, but it can never create or destroy energy; all energy in the forward model is accounted for by radioactive decay.
We apply this framework to the local ZTF sample ($z \leq 0.06$), similar to \citet{Sarin2026b}. However, to prevent the SED from absorbing bolometric shape errors during training, we first learn the spectral surface on the Carnegie Supernova Project (CSP) SN~Ia sample~\citep{Krisciunas2017}, which captures far more of the SED than the more limited ZTF photometry. We then apply this frozen SED model to ZTF, infer the hierarchical physical population and its host dependence, and construct a Tripp-like standardisation from physical and spectral parameters inferred without distance information.

This paper is structured as follows. In Sec.~\ref{sec:model} we describe the bolometric physics model, the learned energy-normalised SED, and the dust and population hierarchy, and in Sec.~\ref{sec:inference} we detail the Monte Carlo stochastic-approximation expectation maximisation (SAEM) inference and physical standardisation procedure. We validate the framework on synthetic populations and radiative-transfer benchmarks in Sec.~\ref{sec:validation}. In Sec.~\ref{sec:application} we apply the model to CSP and ZTF, presenting the inferred physical populations, their host-galaxy trends, the physical standardisation, distance-leakage tests, and the learned SED and its training uncertainty. We discuss the implications and limitations of our results in Sec.~\ref{sec:discussion} and conclude in Sec.~\ref{sec:conclusions}.

\section{Model}
\label{sec:model}

Modelling SN~Ia light curves physically requires two ingredients: a radioactive-decay and radiation-diffusion engine that predicts the total (bolometric) luminosity $\Lbol_i(t)$ escaping the ejecta as a function of time, and a spectral energy distribution (SED) that specifies how that energy is distributed across wavelength so synthetic fluxes can be integrated through observed telescope filters. In a SN~Ia, the brightness is controlled primarily by the synthesised $^{56}$Ni mass $\Mni$, while the rise and post-peak decline are set by the ejecta mass $\Mej$, kinetic energy $\Ek$, and radial $^{56}$Ni distribution through the optical diffusion and gamma-ray escape timescales. Empirical cosmology models such as SALT3 bypass this physics and learn optical flux surfaces directly, whereas purely analytic physical models often assume a simple blackbody SED that fails to match real SN~Ia spectra. 
Our generative model combines both ideas across four coupled components for each SN $i$: a physical bolometric light curve $\Lbol_i(t)$, a learned energy-normalised intrinsic SED that distributes $\Lbol_i(t)$ across wavelength without creating or destroying energy, host-galaxy line-of-sight dust attenuation, and a luminosity distance. 
Host extinction is applied only after the intrinsic SED is normalised, so absorbed energy leaves the observed flux, whereas the intrinsic SED itself never adds or removes bolometric power.

\subsection{Bolometric physics}
\label{sec:snmix}

We compute $\Lbol_i(t)$ using \snmix{}~\citep{Sarin2026a}, a differentiable model of homologously expanding, radioactively heated ejecta implemented in \program{JAX}~\citep{Bradbury2018}. 
The ejecta are discretised into 100 radial shells following a broken power-law density or exponential density profile. 
Each shell is heated by the $^{56}$Ni$\rightarrow{}^{56}$Co$\rightarrow{}^{56}$Fe decay chain, with gamma-ray deposition evaluated from the direction-averaged column density at a fixed gamma-ray opacity of $\kappa_\gamma = 0.03\,{\rm cm^2\,g^{-1}}$ and positron kinetic energy deposited locally. 
The $^{56}$Ni is distributed radially according to a smooth sigmoid profile in enclosed mass fraction $m$, $w(m) = [1 + \exp((m - f_{\rm mix})/0.1)]^{-1}$, normalised to the total $^{56}$Ni mass while keeping the local $^{56}$Ni mass fraction below unity in every shell. 
Thermal energy diffuses between shells and escapes from the surface via an implicit finite-volume solver with a grey optical opacity that transitions smoothly from $0.1$ to $0.03\,{\rm cm^2\,g^{-1}}$ around $22$\,d after explosion. We have verified this coupled-diffusion transport against both the exact integral energy-conservation relation of \citet{Katz2013} and the peak-time--luminosity scaling for mixed radioactive sources of \citet{Khatami2019}.

Each SN is described by five physical parameters: the ejecta mass $\Mej$ (bounded to $0.5$--$2.5\,\Msun$), the $^{56}$Ni mass $\Mni$ (capped at $0.95\,\Mej$), the $^{56}$Ni mixing extent $f_{\rm mix}$ (which early-time light curves and spectral modelling suggest is often substantial; \citealt{Piro2014,Magee2024,Liu2026, Sarin2026a}), the explosion epoch, and the gamma-ray escape column $\propto \Mej^2/\Ek$, which the post-peak decline constrains directly. 
At our fixed opacities, the kinetic energy $\Ek$ and the diffusion timescale follow deterministically from these parameters. Wherever a near-maximum Si\,\textsc{ii}\,$\lambda6355$ velocity $v_{\rm Si}$ is available, we link it to the characteristic bulk ejecta velocity through a log-normal likelihood term,
\be
\ln\sqrt{\frac{10\,\Ek}{3\,\Mej}} \sim \mathcal{N}\!\left(\ln v_{\rm Si},\,
\sqrt{(\sigma_v/v_{\rm Si})^2 + 0.15^2}\right),
\label{eq:velocity}
\ee
where $\sigma_v$ is the quoted measurement uncertainty and the $0.15$ floor accounts for the scatter between a photospheric line velocity and the bulk ejecta speed. Events without a velocity measurement receive a population prior centred at $11\,100\,{\rm km\,s^{-1}}$ with an $18\%$ fractional width. As we demonstrate in Sec.~\ref{sec:populations}, this velocity constraint is often critical: photometry alone constrains $\Mej^2/\Ek$ and allows the fit to trade higher velocity for unphysically heavy ejecta at virtually no cost to the light-curve fit.

\subsection{Learned, energy-normalised SED}
\label{sec:sed}

Following the hierarchical optical-to-near-infrared SED construction of BayeSN~\citep{Mandel2022}, which warps the \citet{Hsiao2007} spectral template with population-level variation modes and per-event residual perturbations, we construct the central intrinsic SED surface of event $i$ at rest-frame phase $t$ around the \citet{Hsiao2007} template $q_{\rm H}(\lambda, t)$ as
\begin{align}
\ln q_i(\lambda, t) =\;& \ln q_{\rm H}(\lambda, t) + \Delta\mu(\lambda, t)
+ \sum_{k=1}^{3} z_{ik}\,\phi_k(\lambda, t) \nonumber\\
&+ \sigma_\epsilon \sum_{r=1}^{30} \epsilon_{ir}\,\psi_r(\lambda, t).
\label{eq:sed}
\end{align}
Each term has a distinct role:
\begin{itemize}
\item $\Delta\mu(\lambda, t)$ is the learned population-mean log-flux deviation from the Hsiao template, shared by all SNe~Ia;
\item $\phi_k(\lambda, t)$ ($k = 1, 2, 3$) are three global spectral variation surfaces shared across the population, weighted for each SN by per-event scores $z_{ik} = \bm{\beta}_k^{\top}\bm{\theta}_i + \tilde{z}_{ik}$, where $\bm{\beta}_k^{\top}\bm{\theta}_i$ captures linear spectral trends with the event's physical parameters $\bm{\theta}_i$ and $\tilde{z}_{ik}\sim\mathcal{N}(0,1)$ allows intrinsic spectral diversity at fixed explosion physics;
\item $\psi_r(\lambda, t)$ ($r = 1,\dots,30$) is a fixed coarse basis of unit spline surfaces ($5$ phase by $6$ wavelength knots), zero-centred at every phase so they only redistribute flux, with standard-normal per-event scores $\epsilon_{ir}\sim\mathcal{N}(0,1)$ scaled by a single shared isotropic scalar $\sigma_\epsilon$. Unlike BayeSN's full-covariance residual surface, this isotropic perturbation is tightly constrained: after training it shrinks to $\lesssim 0.005$\,mag rms at peak, absorbing minor band-by-band residuals without pulling the physical parameters (turning it off changes our $(B-V)_0$ standardisation scatter in Sec.~\ref{sec:standardisation} by $<0.001$\,mag).
\end{itemize}
Both $\Delta\mu$ and $\phi_k$ are tensor-product cubic B-splines with 12 uniform basis functions in phase over $-15$ to $+45$\,d and 14 uniform basis functions in $\ln\lambda$ over $2500$--$25\,000$\,\AA, regularised towards zero so that the Hsiao template acts as the prior mean and the surface departs from it only where it is supported by data. Outside $3300$--$18\,000$\,\AA, where CSP has little or no passband coverage, we restrict the learned deviation. Without this restriction, the spline can shift flux into unobserved far-ultraviolet or far-infrared wavelengths, creating an artificial degeneracy with $\Lbol$.

To complete the full rest-frame wavelength domain ($100$--$300\,000$\,\AA), we combine the learned surface with an ultraviolet component and a near-infrared component whose shapes follow the Hsiao template within its $1000$--$25\,000$\,\AA\ support and transition to smooth, strongly damped log-quadratic continuations beyond it. 
Two per-event parameters control the relative weights of these ultraviolet and near-infrared components against the central surface. 
At every phase, the combined spectrum is normalised to unit integral,
\be
L_{\lambda,i}(\lambda, t) = \Lbol_i(t)\,\frac{q_i(\lambda, t)}{\int q_i(\lambda', t)\,{\rm d}\lambda'},
\label{eq:normalisation}
\ee
where $L_{\lambda,i}$ is the intrinsic specific luminosity and the integral spans $100$--$300\,000$\,\AA. This normalisation guarantees that the SED only redistributes the \snmix{} bolometric luminosity. However, fixing the total energy, does not fix the fraction falling inside a finite set of observed filters, so an unconstrained SED can still absorb bolometric shape errors within the observed window. This is sensitive to how well the data separate physics from spectral evolution, and depends on wavelength coverage. To combat this, we train the SED with each training event's explosion parameters held fixed at standalone quasi-bolometric fits (Sec.~\ref{sec:inference}) and test the bolometric engine directly against radiative-transfer simulations (Sec.~\ref{sec:standart}). 
In addition to the residual spectral surface, the photometric likelihood includes a shared fractional model jitter added in quadrature to the measurement errors.

\subsection{Dust, distance, and population hierarchy}
\label{sec:population}

Host-galaxy line-of-sight extinction is modelled with the \citet{Cardelli1989} reddening law using an exponential population prior, $A_{V,i}\sim{\rm Exponential}(\tau)$, and a single common host total-to-selective extinction ratio $R_V$ learned on CSP (yielding $R_V \simeq 1.85$, which is then held fixed when fitting ZTF). Milky Way foreground extinction is treated separately and fixed to the dust maps of \citet{Schlegel1998} recalibrated by \citet{Schlafly2011} with the standard Galactic $R_V = 3.1$. Each SN's distance modulus is marginalised around its CMB-frame redshift distance,
\be
\mu_i = \mu(z_i) + \sigma_{\mu,i}\,u_i,\qquad u_i\sim\mathcal{N}(0,1),
\label{eq:distance}
\ee
where $\mu(z_i)$ assumes $H_0 = 70\,{\rm km\,s^{-1}\,Mpc^{-1}}$ and $\sigma_{\mu,i}$ combines the redshift measurement uncertainty with $250\,{\rm km\,s^{-1}}$ of peculiar-velocity dispersion in quadrature. During the CSP SAEM inference (and when used as the frozen prior for ZTF), the seven active physical and wavelength-component parameters follow independent Gaussian priors $\mathcal{N}(\mu_k, \sigma_k^2)$ whose means and widths are learned, while the three SED scores $\tilde{z}_{ik}$, the 30 residual scores $\epsilon_{ir}$, and the distance offsets $u_i$ carry standard-normal priors. 
ZTF event posteriors are obtained under a fixed, diagonal CSP prior, therefore individual event estimates scatter with measurement uncertainty and shrink towards the CSP mean. To infer the underlying ZTF physical population and its correlations without refitting every light curve, we perform a hierarchical population analysis in a second step, fitting a full-covariance multivariate Gaussian population (with host-dependent means and separate dust scales) by reweighting each event's posterior draws from the fixed prior~\citep{Hogg2010}.

\section{Inference}
\label{sec:inference}

We find point estimates for the shared global parameters (the SED surface, population hyperparameters, and host dust law) while sampling each SN's individual parameters with Hamiltonian Monte Carlo (HMC). Joint training uses Monte Carlo SAEM~\citep{Delyon1999,Kuhn2004}: each iteration updates the per-event HMC chains at the current global parameters and then steps those global parameters along the gradient of the population likelihood. The pipeline runs in three stages.

\paragraph*{Stage 1: SED training.} Training the SED while explosion parameters float freely would allow the spectral surface to absorb bolometric shape errors. We therefore construct direct quasi-bolometric light curves over $-10$ to $+40$\,d from CSP $u$--$H$ photometry (interpolating well-sampled bands onto a 1-day grid, integrating with the trapezoid rule, and applying a $10$--$20\%$ \citealt{Hsiao2007} correction for unobserved ultraviolet and far-infrared flux). Requiring at least 15 valid epochs selects 46 of the 79 normal CSP SNe; these match the remaining 33 in redshift and colour-stretch $s_{BV}$, with more complete near-infrared coverage. We fit \snmix{} independently to each quasi-bolometric curve at $\mu(z_i)$, hold those explosion parameters fixed, and train the shared SED on broadband CSP photometry for 4500 SAEM iterations (learning rate $3\times10^{-3}$), averaging the final 1800 steps before freezing the SED.

\paragraph*{Stage 2: CSP population.} With the SED frozen, we jointly fit all 79 normal CSP SNe alongside the population means and widths, dust scale $\tau$, and common host $R_V$ (recovering $R_V = 1.85$). Each SN is sampled with 12 HMC chains using a per-event Hessian metric initialised from a grid scan in explosion epoch and ejecta mass (400 warm-up and 2500 production draws per chain). Across the 79 events, median split-$\hat{R}$ lies between $1.003$ and $1.019$ ($\Mej$, $\Mni$, $A_V$, and distance offsets), with $61$--$75/79$ below $1.05$ and $73$--$78/79$ below $1.1$.

\paragraph*{Stage 3: ZTF inference.} Freezing the CSP-trained SED, host $R_V = 1.85$, and CSP population as a fixed prior, we sample each of the 845 ZTF SNe independently with the same 12-chain HMC configuration (achieving median split-$\hat{R} \le 1.01$ across $\Mej$, $\Mni$, and $A_V$, with $740/845$ ejecta masses below $1.05$ and $817/845$ below $1.1$). 
We run two versions of the analysis: a standard redshift-distance run ($\sigma_{\mu,i}$ from redshift and peculiar-velocity errors) for the physical population analysis, and a \emph{redshift-blind} run with $\sigma_{\mu,i} = 3$\,mag so that observed brightness cannot inform the light-curve shape, spectral, or dust parameters. 
All standardisation quantities in Sec.~\ref{sec:standardisation} come exclusively from this single redshift-blind run.

\paragraph*{Physical standardisation.} By analogy with the Tripp relation~\citep{Tripp1998}, we standardise each SN's peak rest-frame $B$ magnitude $m_{B,i}$, evaluated from the redshift-blind forward model (including fitted host dust) through the rest-frame CSP $B$ band between $-10$ and $+15$\,d. Like SALT3 $m_B$, $m_{B,i}$ is independent of distance and retains host extinction. From zero-mean, unit-variance redshift-blind posterior medians $\bm{x}_i$ of chosen physical or spectral predictors (Table~\ref{tab:tripp}), we form the Hubble residual
\begin{equation}
\Delta_i = m_{B,i} - \mu_{{\rm ref},i} - \left(M_0 + \bm{\gamma}^{\top}\bm{x}_i\right),
\label{eq:hubble_residual}
\end{equation}
where $\mu_{{\rm ref},i} \equiv \mu(z_i)$ is the CMB-frame redshift distance modulus, $M_0$ is the mean absolute peak $B$ magnitude, and $\bm{\gamma}$ is the vector of linear standardisation slopes
for the predictors $\bm{x}_i$ (directly analogous to $(-\alpha, \beta)$ in the SALT3 Tripp relation). We fit $M_0$ and $\bm{\gamma}$ by inverse-variance weighted least squares with $\sigma_i^2 =
\sigma_{\mu,i}^2 + \sigma_{m_B,i}^2$. As our physical models span one to four linear predictors compared with two for SALT3, we apply five-fold cross-validation (fitting $M_0$ and $\bm{\gamma}$ on
four fifths of the sample and evaluating on the held-out fifth) to both our models and SALT3 on identical folds, preventing overfitting. 
We report the weighted rms (${\rm wRMS} = [\sum_i w_i \Delta_i^2 / \sum_i w_i]^{1/2}$), the outlier-resistant normalised median absolute deviation (${\rm NMAD} = 1.4826 \times {\rm median}(|\Delta_i - {\rm median}(\Delta)|)$), and the intrinsic scatter $\sigma_{\rm int}$. Following standard SN~Ia cosmology practice, the host-mass step (across $\log_{10}(M_*/\Msun) = 10$) and host-colour step (across median rest-frame host $g-z = 1.086$\,mag) are the weighted mean residual differences with weights $w_i = (\sigma_i^2 + \sigma_{\rm int}^2)^{-1}$, where $\sigma_{\rm int}$ is fitted simultaneously so that the reduced $\chi^2$ around the two host-bin means equals unity.

\section{Validation}
\label{sec:validation}

\subsection{Recovery of synthetic populations}
\label{sec:synthetic}

To test pipeline calibration, we simulated three 80-event CSP-like samples from the generative model with real CSP cadences, bands, and flux errors, and fitted each with the full sampling procedure. Across all three seeds, $\Mni$ and distance modulus achieve well-calibrated $95\%$ credible-interval coverage ($0.95$ and $0.96$).  
Ejecta mass and explosion epoch are unbiased on average with intervals $20$--$30\%$ narrower than nominal ($95\%$ coverage $0.88$), while $A_V$ is biased low by $\simeq 0.25\sigma$ ($0.01$--$0.03$\,mag). Including the Si\,\textsc{ii} velocity likelihood (equation~\ref{eq:velocity}) breaks the velocity--mass degeneracy, improves ejecta-mass convergence, and resolves the $A_{V}$ bias.

We also tested host-dependence recovery on 150 synthetic ZTF-cadence events (103 in high-mass hosts), injecting host-mass shifts into the gamma-escape column, $^{56}$Ni parameter, and dust scale $\tau$ while setting the true shifts in $\Mej$ and mixing to zero. Hierarchical reweighting recovers the injected decline, $^{56}$Ni, and dust shifts within $\simeq 1\sigma$ and leaves $\Mej$ consistent with zero ($0.07\sigma$), but produces a spurious $+1.2\sigma$ shift in mixing; we therefore do not draw strong conclusions from small host-dependent mixing shifts alone.

\subsection{Comparison with radiative-transfer models}
\label{sec:standart}

\begin{figure*}
\centering
\includegraphics[width=\textwidth]{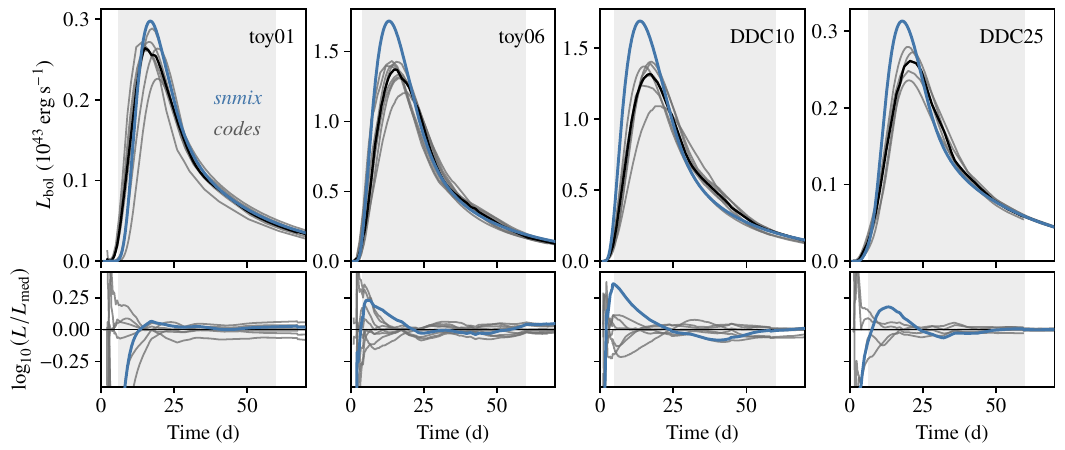}
\caption{Comparison of \snmix{} against the StaNdaRT radiative-transfer benchmarks~\citep{Blondin2022}. Top panels: bolometric light curves of the individual radiative-transfer codes (grey), their median (black), and \snmix{} evaluated on each benchmark's exact density and $^{56}$Ni profile with our adopted opacity law and the true $^{56}$Ni mass (blue), shown for toy01, toy06, DDC10, and DDC25 from left to right. Bottom panels: fractional residuals of each curve relative to the code median. The shaded band marks the $3$--$60$\,d fitting window. Around peak, \snmix{} is systematically brighter than every code ($13$--$28\%$ above the median and $3$--$20\%$ above the brightest code) while matching the post-peak tail, so whole-light-curve fits recover $^{56}$Ni to within $2$--$6\%$ whereas peak-only fits would bias $^{56}$Ni low.}
\label{fig:standart}
\end{figure*}

As \snmix{} uses grey shell-by-shell opacities (relying on the learned SED for wavelength distribution), we benchmark its bolometric light curves against full multi-group radiative-transfer simulations from the StaNdaRT project~\citep{Blondin2022}. StaNdaRT evolved four standardised Type~Ia ejecta structures (toy01, toy06, DDC10, and DDC25) across independent codes. Supplying each benchmark's exact density and $^{56}$Ni profile to \snmix{}, we fit the codes' median bolometric light curve over $3$--$60$\,d (above $10\%$ of peak luminosity) with only $\Mni$ free. Figure~\ref{fig:standart} compares \snmix{} evaluated at each model's true $^{56}$Ni mass against the radiative-transfer codes.

Across all four benchmarks, \snmix{} recovers $\Mni$ to within $2$--$6\%$ (Table~\ref{tab:standart}), well inside the $11$--$27\%$ inter-code spread, with overall shape agreement of $0.07$--$0.19$\,dex rms. However, at the true $^{56}$Ni mass \snmix{} is systematically brighter around peak (days $5$--$20$), sitting $13$--$28\%$ above the code median ($3$--$20\%$ above the brightest code), and peaks $2$--$3$\,d earlier in three of four models. Because the post-peak tail matches or sits slightly below the code median, whole-light-curve fits recover $\Mni$ accurately whereas peak-only fits would underestimate $\Mni$ by $11$--$22\%$. In real data, an early model peak can also bias absolute $\Mej$ upward to lengthen the diffusion time. Testing a temperature-dependent opacity law (held out across benchmarks) improves the light-curve rms on three of four models without changing the recovered $^{56}$Ni masses (Sec.~\ref{sec:nickel_distance}), confirming that our grey opacity primarily affects peak shape rather than the $^{56}$Ni mass.

\begin{table}
\centering
\small
\caption{Validation of \snmix{} against the four StaNdaRT radiative-transfer benchmarks~\citep{Blondin2022}. The rms is evaluated in $\log_{10}\Lbol$ relative to the codes' median bolometric light curve over $3$--$60$\,d (above $10\%$ of peak). The $\Mni$ ratio gives the median over codes of the recovered-to-true $^{56}$Ni mass using the opacity prescription of this paper, and the code spread gives the full range of that ratio across codes. For every benchmark, the recovered $^{56}$Ni mass falls well within the inter-code spread.}
\label{tab:standart}
\begin{tabular}{lccc}
\toprule
Model & rms (dex) & $\Mni$ ratio & Code spread \\
\midrule
toy01 & 0.19 & 1.03 & 0.27 \\
toy06 & 0.08 & 0.96 & 0.14 \\
DDC10 & 0.12 & 0.94 & 0.14 \\
DDC25 & 0.07 & 0.98 & 0.11 \\
\bottomrule
\end{tabular}
\end{table}

\section{Application to CSP and ZTF}
\label{sec:application}

\subsection{Data}
\label{sec:data}

We use two samples across our study. Table~\ref{tab:sample} summarises the sample selection for both surveys.

\paragraph*{Carnegie Supernova Project.} We use optical and near-infrared natural-system photometry from CSP DR3~\citep{Krisciunas2017} over rest-frame phases $-15$ to $+45$\,d. Restricting to spectroscopically normal SNe~Ia with colour-stretch $s_{BV}\ge 0.8$~\citep{Burns2014} (excluding 16 fast decliners and one strong outlier, SN~2006dd) leaves 79 events at $z = 0.005$--$0.084$ (median $0.023$). We exclude $s_{BV}<0.8$ events because our SED cannot reproduce their intrinsically redder colours, causing joint fits to assign them excess host extinction ($\Delta E(B-V)\simeq 0.10$--$0.19$\,mag above \citealt{Burns2018}) or unphysical $^{56}$Ni masses. As in Sec.~\ref{sec:inference}, 46 events train the SED, 14 low-reddening calibrators have host extinction held near \citet{Burns2018} to break the dust--SED zero-point degeneracy, and 50 events have near-maximum Si\,\textsc{ii}\,$\lambda6355$ velocities from \citet{Folatelli2013}.

\paragraph*{Zwicky Transient Facility.} From ZTF DR2~\citep{Rigault2025a}, we select 845 spectroscopically normal SNe~Ia at $z<0.06$ (median $0.044$) with $g$, $r$, and $i$ photometry over $-10$ to $+40$\,d, removing 268 of the 1113 normal events that fail light-curve coverage requirements (Table~\ref{tab:sample}). The dropped events show no statistically significant difference in SALT3 $x_1$ ($-0.27\pm0.11$ vs.\ $-0.39\pm0.04$), colour, or host mass, and are nominally considered a volume-complete sample in the $z<0.06$ regime (excluding peculiar subtypes and unclassified events). Although, we stress that completeness to SALT3 does not imply true completeness. Coverage requirements can introduce selection biases, potentially including brightness-dependent selection~\citep{Sarin2026b}. 

Near-maximum Si\,\textsc{ii} velocities are available for 384 ZTF SNe~\citep{Burgaz2025a} (median $11\,100\,{\rm km\,s^{-1}}$, which we adopt as the prior for events without spectroscopic measurements). 
Global host masses and rest-frame $g-z$ colours are available for 837 events, and local host properties for 718 of the 727 standardisation events. 
We adopt global host properties as our baseline to match standard SN~Ia cosmology; while local host steps are larger in some surveys~\citeg{Rigault2013,Rigault2020,Roman2018,Kelsey2021,Kelsey2023,Kelsey2026}, they are comparable in ZTF, and we verify both in Sec.~\ref{sec:standardisation}. 
Applying standard cosmology cuts ($|x_1|<3$, $|c|<0.3$, accepted SALT3 fit, measured host mass) yields 727 events (390 above $10^{10}\,\Msun$). 
We refit all light curves with SALT3~\citep{Kenworthy2021} using \program{jax-SNANA}~\citep{Grayling2026} and \texttt{SALT3.DOVEKIE}~\citep{Popovic2026}.

\paragraph*{Distances.} Each SN's distance modulus is marginalised around its CMB-frame redshift distance (equation~\ref{eq:distance}) at fixed $H_0 = 70\,{\rm km\,s^{-1}\,Mpc^{-1}}$ (or $\sigma_{\mu,i} = 3$\,mag in the redshift-blind run). Without a temperature-dependent opacity, distance and $\Mni$ are strictly degenerate, so absolute $^{56}$Ni masses scale as $(H_0/70\,{\rm km\,s^{-1}\,Mpc^{-1}})^{-2}$ while differential host and standardisation comparisons are independent of $H_0$.

\begin{table}
\centering
\small
\caption{Summary of sample selection for CSP DR3 and ZTF DR2.}
\label{tab:sample}
\begin{tabular}{llr}
\toprule
Survey & Selection criterion & Events \\
\midrule
CSP DR3 & All white-dwarf explosions & 134 \\
 & Primary Type~Ia / spectroscopically normal & 123 / 96 \\
 & Normal with $s_{BV}\ge 0.8$ (excl.\ SN~2006dd) & 79 \\
 & Quasi-bolometric subsample (SED training) & 46 \\
 & Measured Si\,\textsc{ii} velocity~\citep{Folatelli2013} & 50 \\
\midrule
ZTF DR2 & All SNe~Ia at $z<0.06$ & 1583 \\
 & Spectroscopically normal (incl.\ 04gs-like) & 1113 \\
 & Coverage flag and $\ge 3$ points in $g$ and $r$ & 845 \\
 & Global host mass and rest-frame $g-z$ & 837 \\
 & Measured Si\,\textsc{ii} velocity~\citep{Burgaz2025a} & 384 \\
 & Standardisation sample ($|x_1|<3$, $|c|<0.3$, host) & 727 \\
\bottomrule
\end{tabular}
\end{table}

\subsection{Physical populations}
\label{sec:populations}

\begin{figure}
\centering
\includegraphics[width=\columnwidth]{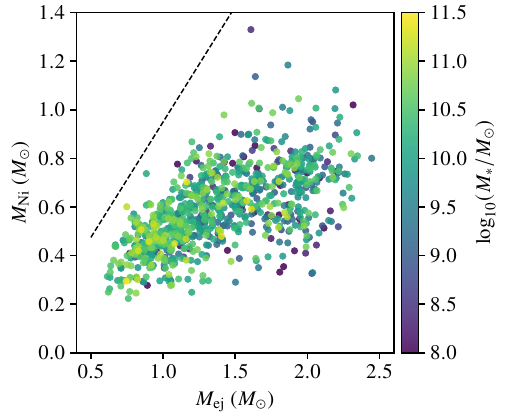}
\caption{Inferred $^{56}$Ni mass against ejecta mass for the 837 ZTF SNe with measured host masses, coloured by host stellar mass. Points show per-event posterior medians from the redshift-distance fit, and the dashed line indicates the physical boundary $\Mni = 0.95\,\Mej$. The $^{56}$Ni and ejecta masses correlate strongly ($r = 0.61$), with SNe in massive hosts preferentially populating lower ejecta masses.}
\label{fig:population}
\end{figure}

We summarise the per-event posterior medians for both surveys in Table~\ref{tab:pops} and display the ZTF events in the $\Mni$--$\Mej$ plane in Fig.~\ref{fig:population}. CSP and ZTF agree closely on the $^{56}$Ni mass scale, with per-event medians of $0.60$ and $0.57\,\Msun$, respectively. ZTF SNe have slightly lighter median ejecta ($1.33$ against $1.55\,\Msun$) and lower host extinction ($A_V = 0.25$ against $0.29$\,mag). ZTF is considered complete to cosmological Ia SNe within the $z \leq 0.06$ volume, and so these differences between CSP and ZTF could reflect that fact. We again stress that ZTF survey completeness is predicated on survey efficiency against a SALT model, which implies little about SNe~Ia that may not be well described by a SALT template. Across ZTF, $^{56}$Ni and ejecta mass are strongly correlated (Pearson $r = 0.61$), and no SN sits against the model's $\Mni\le 0.95\,\Mej$ ceiling.

One of the strongest drivers of our mass constraints is the Si\,\textsc{ii} velocity likelihood (equation~\ref{eq:velocity}). Without it, the post-peak light curve fixes only the gamma-ray escape column $\propto\Mej^2/\Ek$, allowing the sampler to inflate $\Mej$ by simultaneously raising the ejecta speed. In an unconstrained CSP fit, 35 of the 79 events exceed $\Mej = 2.1\,\Msun$ and imply bulk velocities of $\simeq 14\,800\,{\rm km\,s^{-1}}$, roughly $4000\,{\rm km\,s^{-1}}$ higher than their measured Si\,\textsc{ii} lines ($\simeq 10\,800\,{\rm km\,s^{-1}}$). Introducing equation~(\ref{eq:velocity}) reduces the count above $2.1\,\Msun$ from 35 to just 2 events while worsening the photometric fit by only $\simeq 1$ unit of $\chi^2$ per supernova across a mean of 123 photometric observations per CSP event ($<0.01$ in $\chi^2$ per observation). This is a by-product of the well-known ejecta mass--diffusion time degeneracy, which can be broken by additional prior constraints. With the velocity term, the implied CSP bulk speeds are $\simeq 10\%$ higher than the Si\,\textsc{ii} line velocities ($12\,160$ against $10\,990\,{\rm km\,s^{-1}}$), which is within the expected offset between a line-forming layer and a model's characteristic ejecta velocity.

\begin{table}
\centering
\small
\caption{Summary of per-event posterior medians for CSP and ZTF, with 16th--84th percentiles across events shown in parentheses for $\Mej$. These values summarise the fitted event medians rather than the hierarchical population widths. The two surveys agree closely on $^{56}$Ni mass, while the untargeted ZTF sample shows slightly lighter ejecta and less dust.}
\label{tab:pops}
\begin{tabular}{lcc}
\toprule
 & CSP (79) & ZTF (845) \\
\midrule
$\Mej$ ($\Msun$) & 1.55 (0.98--1.87) & 1.33 (0.96--1.92) \\
$\Mni$ ($\Msun$) & 0.60 & 0.57 \\
$A_V$ (mag) & 0.29 & 0.25 \\
\bottomrule
\end{tabular}
\end{table}

Table~\ref{tab:pops} summarises individual event medians under the fixed CSP prior. Hierarchical population inference across the 837 events with host masses (reweighting each event's posterior draws to a full-covariance Gaussian population with host-dependent means and dust scales; \citealt{Hogg2010}) yields: $\Mej = 1.39\,(1.02\text{--}1.72)\,\Msun$, $\Mni = 0.57\,(0.43\text{--}0.71)\,\Msun$, and burning efficiency $\Mni/\Mej = 0.42\,(0.34\text{--}0.50)$ ($<0.1\%$ above $0.8$). By mass bin, $32\%$ lie below $1.2\,\Msun$, $30\%$ between $1.2$ and $1.5\,\Msun$, and $38\%$ above $1.5\,\Msun$. The light curves also strongly constrain $^{56}$Ni mixing (shrinking posteriors to roughly one third of the prior width), favouring extensive mixing ($f_{\rm mix} = 0.59$, $16$th--$84$th percentiles $0.47$--$0.73$), consistent with shallow early rises in ZTF~\citep{Liu2026}. Using the resulting hierarchical per-event means, $\Mej$ correlates with SALT3 $x_1$ at $r = 0.81$, $\Mni$ with $x_1$ at $r = 0.70$, and $A_V$ with $c$ at $r = 0.90$ (Spearman $\rho = 0.85$), with no significant mass or dust difference between high- and normal-velocity SNe ($v_{\rm Si} > 12\,000\,{\rm km\,s^{-1}}$). Unlike~\citet{Sarin2026b}, our hierarchical analysis ignores any selection term and therefore, only applies to our selected ZTF sample and should not be used to interpret the broader population of Type Ia SNe without caution. 

\begin{figure*}
\centering
\includegraphics[width=\textwidth]{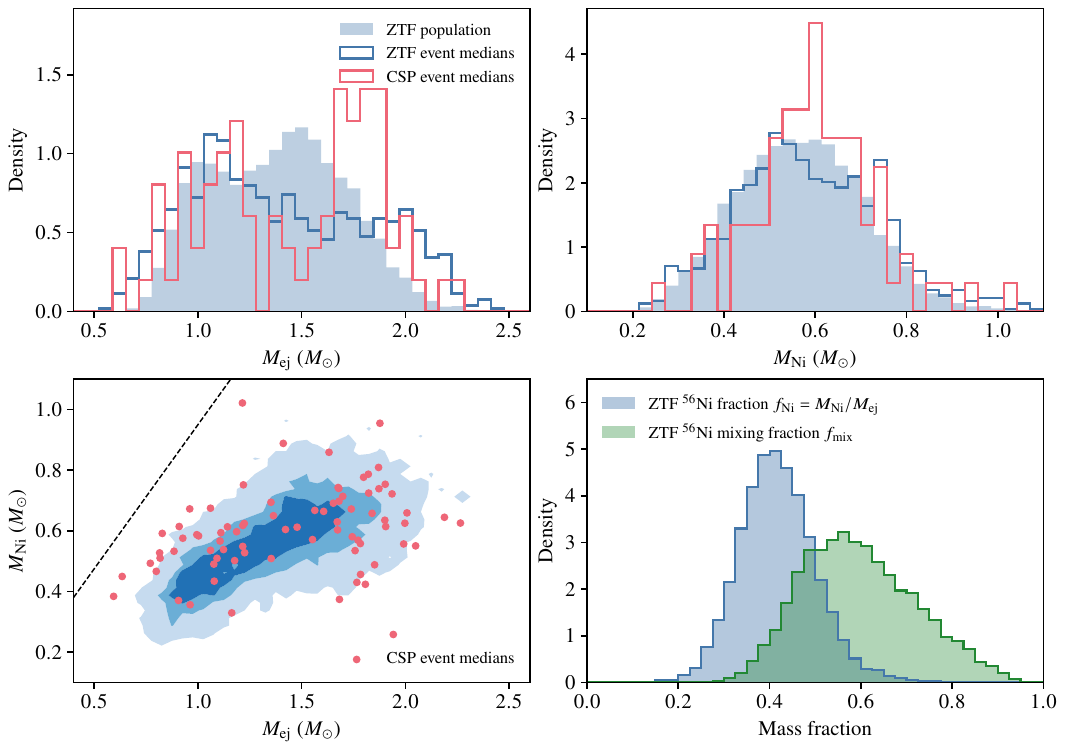}
\caption{Hierarchical physical population of the 837 ZTF SNe with host masses compared against CSP. Top left: ejecta mass; top right: $^{56}$Ni mass, showing the hierarchical ZTF population (filled histogram, inferred by reweighting each event's posterior draws), raw ZTF event posterior medians (blue line), and CSP event posterior medians (79 events with the Si\,\textsc{ii} velocity constraint, rose line). Bottom left: joint $^{56}$Ni--$\Mej$ distribution, where shaded contours enclose $39$, $68$, and $95\%$ of the hierarchical ZTF population, rose points mark CSP event medians, and the dashed line shows $\Mni = 0.95\,\Mej$. Bottom right: hierarchical ZTF distributions of the $^{56}$Ni mass fraction ($f_{\rm Ni} = \Mni/\Mej$, blue) and the $^{56}$Ni mixing mass fraction ($f_{\rm mix}$, green).}
\label{fig:popdist}
\end{figure*}

In Figure~\ref{fig:popdist}, we compare this ZTF population with raw ZTF and CSP event medians. The hierarchical analysis narrows the $\Mej$ spread from $0.95$--$1.93\,\Msun$ to $1.02$--$1.72\,\Msun$ by accounting for measurement uncertainty along the mixing--ejecta-mass degeneracy, while leaving $\Mni$ nearly unchanged because brightness at the redshift distance pins down $^{56}$Ni tightly given the empirical SED prior anchored on the Hsiao template. CSP events sit squarely within the ZTF population contours.

In five-fold cross-validation of the hierarchical ZTF population, a two-component Gaussian mixture improves the held-out log-likelihood over a single Gaussian by $\Delta\ln\mathcal{L} = +88\pm24$ ($\Delta{\rm BIC} = -160$), separating a smaller component ($\sim 25$--$30\%$ of events, $\Mej\simeq 0.95\,\Msun$, $\Mni\simeq 0.46\,\Msun$, $f_{\rm mix}\simeq 0.8$, preferentially in massive hosts) from a dominant component ($\Mej\simeq 1.53\,\Msun$, $f_{\rm mix}\simeq 0.53$). However, this gain comes entirely from the 461 events without measured Si\,\textsc{ii} velocities ($\Delta\ln\mathcal{L} = +51\pm19$) and disappears among the 384 events with velocity measurements ($\Delta\ln\mathcal{L} = -19\pm50$). Because the two components also separate directly along the strong per-event $\Mej$--$f_{\rm mix}$ posterior degeneracy ($r\simeq -0.9$), we do not interpret them as distinct progenitor channels; our distribution is therefore not inconsistent with the unimodal ZTF-BTS $^{56}$Ni distribution of \citet{Sharon2022}.

Separating the learned SED from explosion physics resolves the main biases of one-zone blackbody analysis of ZTF DR2~\citep{Sarin2026b}. While rescaling their truncated hierarchical mean $^{56}$Ni mass ($0.64\pm0.06\,\Msun$ at $H_0 = 67.4\,{\rm km\,s^{-1}\,Mpc^{-1}}$; \citealt{Planck2020}) to $H_0 = 70$ gives $0.59\,\Msun$, the raw per-event medians drop sharply from $\approx 1.2\,\Msun$ to $0.57\,\Msun$ (close to their $0.45\,\Msun$ pseudo-bolometric lower limit), burning efficiency drops from $f_{\rm Ni}\simeq 0.8$--$0.88$ to $0.42$, and the host-mass $^{56}$Ni offset shrinks by more than a factor of three (from $0.13$ to $0.04\pm0.01\,\Msun$) while the $50\%$ host-extinction offset vanishes under a common dust law (Sec.~\ref{sec:host}).

Our bolometric scales also match direct observational bounds. At the hierarchical ZTF population median ($\Mej = 1.39\,\Msun$, $\Ek = 1.19$\,foe, $\Mni = 0.57\,\Msun$, $f_{\rm mix} = 0.59$), \snmix{} peaks at $\Lbol_{\rm peak} = 1.51\,(1.21\text{--}1.80)\times10^{43}\,{\rm erg\,s^{-1}}$ ($\sim 1.3\times10^{43}\,{\rm erg\,s^{-1}}$ accounting for the $\sim 15\%$ peak overshoot against StaNdaRT), inside the observed normal-SN~Ia bound ($\le 1.65\times10^{43}\,{\rm erg\,s^{-1}}$; \citealt{Sharon2025}) and well below the $2.9\times10^{43}\,{\rm erg\,s^{-1}}$ blackbody medians of \citet{Sarin2026b}. The implied uniform-sphere gamma-ray escape time is $t_0 = 43.5\,(37.3\text{--}47.8)$\,d (correlating with $\Mni$ at $r = 0.69$; direct late-time \snmix{} deposition gives $\simeq 45$\,d), sitting just above the $30$--$44$\,d Katz-integral range of \citet{Sharon2020,Sharon2020b,Sharon2025} (which includes fainter fast decliners), consistent with the upward $\Mej$ systematics in Sec.~\ref{sec:limitations}.

\subsection{Host-galaxy dependence on physical parameters}
\label{sec:host}

\begin{figure}
\centering
\includegraphics[width=\columnwidth]{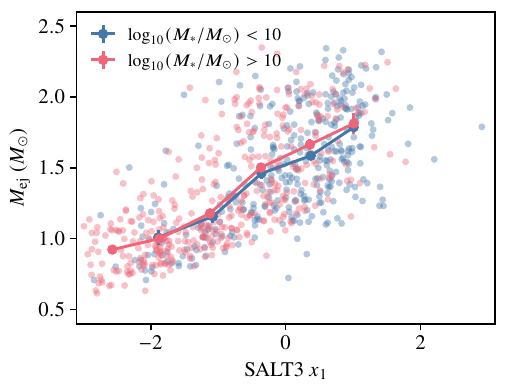}
\caption{Ejecta mass against SALT3 stretch $x_1$ for the 727 ZTF SNe in the standardisation sample, split into low-mass ($\log_{10}(M_*/\Msun)<10$, blue, 337 events) and high-mass (rose, 390 events) hosts. Points show per-event posterior medians and connected markers show bin means with standard errors. Both host populations trace the same $\Mej$--$x_1$ locus ($\simeq 0.26\,\Msun$ per unit $x_1$) and differ only in where they sit along it: the naive $-0.20\,\Msun$ host difference vanishes to $+0.04\,\Msun$ at fixed $x_1$.}
\label{fig:mej_x1}
\end{figure}

Splitting the ZTF sample at $\log_{10}(M_*/\Msun) = 10$, we find that SNe~Ia in high-mass hosts have lighter ejecta by $0.20\pm0.03\,\Msun$, slightly less $^{56}$Ni (by $0.04\pm0.01\,\Msun$), and, under our common dust law, indistinguishable host extinction ($\Delta A_V = +0.02\pm0.02$\,mag). This ejecta-mass offset tracks the well-known stretch--host relation closely, as shown in Fig.~\ref{fig:mej_x1}. On the 727-event standardisation sample, the redshift-distance per-event posterior median ejecta masses correlate tightly with SALT3 $x_1$ ($r = 0.70$, or $0.67$ in $\ln\Mej$, rising to $0.81$ for the hierarchical per-event means), and $x_1$ is lower by $0.93$ in high-mass hosts.

Fitting $\Mej$ jointly against $x_1$ and host-mass bin, the residual host offset drops from $-0.20\pm0.03\,\Msun$ to $+0.04\pm0.03\,\Msun$ ($+0.02\pm0.02\,\Msun$ using the redshift-blind masses), and stays flat at $+0.03\pm0.02\,\Msun$ when $x_1^2$ and $c$ are added. Equivalently, at fixed $x_1$, $\ln\Mej$ shows essentially no residual correlation with continuous host stellar mass ($r = +0.04$). We obtain the same behaviour separately for the 320 events with measured Si\,\textsc{ii} velocities from \citet{Burgaz2025a} (shifting from $-0.27\pm0.05$ to $+0.02\pm0.04\,\Msun$ at fixed $x_1$) and for the 407 events relying on the population velocity prior (shifting from $-0.15\pm0.04$ to $+0.05\pm0.03\,\Msun$). The velocity prior therefore does not manufacture the ejecta-mass trend, and there is no evidence for an additional ejecta-mass offset between host classes at fixed stretch. Conversely, however, SALT3 $x_1$ correlates more strongly with host stellar mass ($r = -0.45$) than $\ln\Mej$ alone ($r = -0.27$): even at fixed $\Mej$, SNe in massive hosts still have lower $x_1$ (residual correlation $r = -0.37$), because empirical optical stretch responds to intrinsic colour and cooling as well as ejecta mass (Sec.~\ref{sec:standardisation}).

\subsection{Physical standardisation}
\label{sec:standardisation}

\begin{figure*}
\centering
\includegraphics[width=\textwidth]{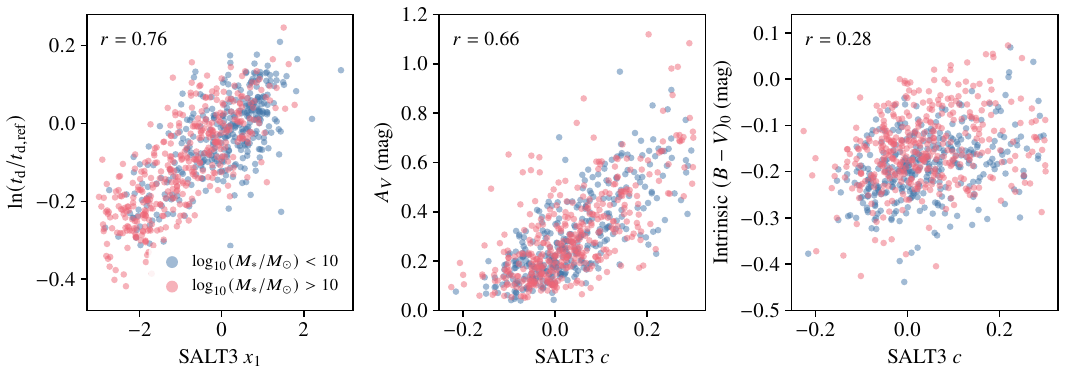}
\caption{Empirical SALT3 parameters expressed in physical terms for the 727 ZTF SNe of Table~\ref{tab:tripp} (redshift-blind posterior medians, coloured by host stellar mass below $10^{10}\,\Msun$ in blue and above in red). Left: diffusion timescale $t_{\rm d}$ against SALT3 $x_1$ ($r = 0.76$). Centre: host extinction $A_V$ against SALT3 $c$ ($r = 0.66$; one highly reddened event lies above the plotted range). Right: peak rest-frame intrinsic $(B-V)_0$ colour, evaluated from each event's dust-free learned SED, against SALT3 $c$ ($r = 0.28$, strengthening to $0.54$ for $A_V<0.1$\,mag; one blue outlier lies below the plotted range), showing that SALT3-redder events are intrinsically redder as well as more extinguished.}
\label{fig:salttranslation}
\end{figure*}

Two timescales govern the bolometric light-curve shape in our physical model, the diffusion timescale $t_{\rm d}\propto\Mej^{3/4}\Ek^{-1/4}$ sets the peak width (correlating with SALT3 $x_1$ at $r = 0.76$ on redshift-blind medians), while the gamma-ray escape timescale $t_\gamma\propto\Mej\Ek^{-1/2}$ sets the post-peak decline (with $t_{\rm d}\propto\Mej^{1/4}t_\gamma^{1/2}$ at fixed opacities). Colour separates into extrinsic host extinction $A_V$ and the SN's intrinsic spectral shape, which we summarise via rest-frame peak intrinsic colours such as $(B-V)_0$ and $(r-i)_0$ integrated from each event's dust-free redshift-blind SED. 
As Fig.~\ref{fig:salttranslation} shows, SALT3 $c$ blends both channels: on redshift-blind medians, $c$ correlates with $A_V$ ($r = 0.66$, rising to $0.90$ for hierarchical redshift-distance means) and with intrinsic $(B-V)_0$ ($r = 0.28$, strengthening to $r = 0.54$ among the 50 low-extinction events with $A_V < 0.1$\,mag). Peak $(B-V)_0$ also anti-correlates with $x_1$ ($r = -0.50$), even though SALT3 $x_1$ and $c$ are weakly correlated in broadband fits: once dust is separated out, narrower, cooler SNe~Ia are intrinsically redder at peak likely due to faster iron-group cooling.

Table~\ref{tab:tripp} and Fig.~\ref{fig:step_sizes} compare a ladder of physical standardisations against SALT3 on the same 727 ZTF SNe. Without standardisation, peak $B$ magnitudes have $0.35$\,mag wRMS ($0.360$\,mag NMAD) and a mild positive host-mass step ($+0.044\pm0.026$\,mag; SALT3 $+0.027\pm0.026$\,mag), reflecting that SNe~Ia in massive hosts are narrower and intrinsically fainter. SALT3 $x_1$ alone flips the step to $-0.083\pm0.024$\,mag ($0.324$\,mag wRMS), whereas standardising on $t_{\rm d}$, $(t_{\rm d}, A_V)$, or $(t_{\rm d}, t_\gamma, A_V)$ lowers scatter to $0.273$--$0.277$\,mag while leaving the host-mass step near zero ($+0.002$ to $+0.007$\,mag) alongside a mild positive host-colour step ($+0.034$ to $+0.040\pm0.020$\,mag; see Sec.~\ref{sec:empirical}).

Adding peak intrinsic colour $(B-V)_0$ collapses the scatter to match or beat SALT3 with three to four interpretable SN properties. Using $(\Mej, A_V, (B-V)_0)$ matches SALT3 in wRMS ($0.177$ vs.\ $0.180$\,mag), NMAD ($0.161$ vs.\ $0.160$\,mag), and $\sigma_{\rm int}$ ($0.151$ vs.\ $0.150$\,mag) while lowering the host-mass step to $-0.073\pm0.013$\,mag (host-colour step $-0.053\pm0.013$\,mag). Replacing $\Mej$ with $t_{\rm d}$ gives $0.175$\,mag wRMS and a $-0.083\pm0.013$\,mag step, and adding $(r-i)_0$ to form $(t_{\rm d}, A_V, (B-V)_0, (r-i)_0)$ beats SALT3 on every scatter metric ($0.173$\,mag wRMS, $0.152$\,mag NMAD, $\sigma_{\rm int} = 0.147$\,mag) with a $-0.088\pm0.013$\,mag step. Across all competitive rows, NMAD sits $\simeq 0.04$\,mag below unweighted rms ($0.198$--$0.202$\,mag), showing similar outlier tails in SALT3 and our physical models. Evaluating $(B-V)_0$ without the 30 per-event residual spline terms changes scatter by $<0.001$\,mag, and using all ten model parameters offers no further gain ($0.177$\,mag wRMS, $-0.072\pm0.013$\,mag step).

\begin{table*}
\centering
\small
\caption{Five-fold cross-validated standardisation of peak rest-frame $B$ magnitudes for the 727 ZTF SNe. Each physical row uses redshift-blind posterior medians of the listed SN properties to predict the redshift-blind model's $m_B - \mu_{\rm ref}$, compared against SALT3 cross-validated on the same events and folds. $N_{\rm par}$ gives the number of linear standardisation parameters. The host-mass step is evaluated across $\log_{10}(M_*/\Msun)=10$ and the host-colour step across the median rest-frame host $g-z$ ($1.086$\,mag); wRMS is inverse-variance weighted and ${\rm NMAD} = 1.4826 \times {\rm median}(|\Delta_i - {\rm median}(\Delta)|)$.}
\label{tab:tripp}
\begin{tabular}{lcccccc}
\toprule
Standardisation parameters & $N_{\rm par}$ & Mass step (mag) & Colour step (mag) & wRMS (mag) & NMAD (mag) & $\sigma_{\rm int}$ (mag) \\
\midrule
SALT3 $x_1$, $c$ & 2 & $-0.093\pm0.014$ & $-0.071\pm0.013$ & 0.180 & 0.160 & 0.150 \\
SALT3 $x_1$ only & 1 & $-0.083\pm0.024$ & $-0.050\pm0.024$ & 0.324 & 0.315 & 0.302 \\
SALT3, none ($m_B-\mu_{\rm ref}$ only) & 0 & $+0.027\pm0.026$ & $+0.080\pm0.026$ & 0.345 & 0.348 & 0.324 \\
\midrule
None ($m_B-\mu_{\rm ref}$ only) & 0 & $+0.044\pm0.026$ & $+0.102\pm0.026$ & 0.350 & 0.360 & 0.332 \\
$t_{\rm d}$ & 1 & $+0.002\pm0.026$ & $+0.049\pm0.026$ & 0.343 & 0.346 & 0.325 \\
$t_{\rm d}$, $A_V$ & 2 & $+0.002\pm0.021$ & $+0.034\pm0.021$ & 0.277 & 0.269 & 0.256 \\
$t_{\rm d}$, $t_\gamma$, $A_V$ & 3 & $+0.007\pm0.020$ & $+0.040\pm0.020$ & 0.273 & 0.261 & 0.252 \\
$\Mej$, $A_V$, $(B-V)_0$ & 3 & $-0.073\pm0.013$ & $-0.053\pm0.013$ & 0.177 & 0.161 & 0.151 \\
$t_{\rm d}$, $A_V$, $(B-V)_0$ & 3 & $-0.083\pm0.013$ & $-0.065\pm0.013$ & 0.175 & 0.161 & 0.149 \\
$t_{\rm d}$, $A_V$, $(B-V)_0$, $(r-i)_0$ & 4 & $-0.088\pm0.013$ & $-0.072\pm0.013$ & 0.173 & 0.152 & 0.147 \\
\bottomrule
\end{tabular}

\vspace{1.0em}
\includegraphics[width=\textwidth]{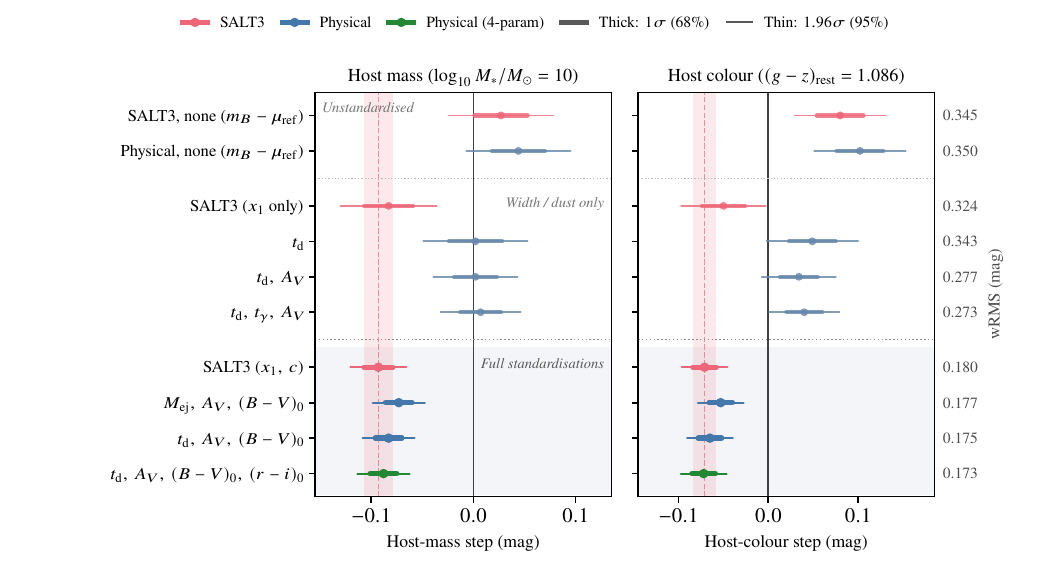}
\captionof{figure}{Host-galaxy step sizes across the standardisation ladder of Table~\ref{tab:tripp} for the 727 ZTF SNe, showing the global host-mass step at $\log_{10}(M_*/\Msun) = 10$ (left) and the global rest-frame host $g-z$ colour step at the sample median of $1.086$\,mag (right), with each model's weighted rms listed along the right edge. Thick bars mark $1\sigma$ ($68\%$) uncertainties and thin whiskers mark $1.96\sigma$ ($95\%$); the vertical shaded rose band marks the $1\sigma$ interval of SALT3 $(x_1, c)$. Standardising on diffusion time and host dust alone lands near zero in host mass while leaving $0.273$--$0.277$\,mag scatter and a mild positive host-colour step, whereas adding peak intrinsic colour $(B-V)_0$ (shaded bottom block) matches or improves upon SALT3 scatter while recovering a $-0.073$ to $-0.088$\,mag step.}
\label{fig:step_sizes}
\end{table*}

Using a paired bootstrap over the 727 identical events (5000 resamples), $(\Mej, A_V, (B-V)_0)$ reduces the host-mass step relative to SALT3 by $0.020$\,mag ($95\%$ interval $0.006$--$0.034$\,mag; $<0.001$\,mag sensitivity across 20 fold partitions) and the host-colour step by $0.018$\,mag ($95\%$ interval $0.003$--$0.032$\,mag), while matching SALT3 in unweighted rms ($0.201$ vs.\ $0.202$\,mag) and lowering the unweighted mass step from $-0.110$ to $-0.089$\,mag. Host mass and colour are distinct proxies for the underlying environmental driver: stellar mass tracks integrated star-formation history and metallicity, whilst colour tracks recent star-formation rate and dust. Consequently, a standardisation that removes the host-mass step need not remove the host-colour step, as seen in the width-and-dust-only rows of Fig.~\ref{fig:step_sizes}. Tightening the scatter with $(t_{\rm d}, A_V, (B-V)_0)$ or $(t_{\rm d}, A_V, (B-V)_0, (r-i)_0)$ moves both steps back closer to SALT3 (paired mass-step reductions $0.010$ and $0.005$\,mag), even as the four-parameter model lowers wRMS below SALT3 by $0.007$\,mag ($95\%$ interval $0.000$--$0.013$\,mag).

These trends hold regardless of split threshold or global versus local host tracers. Although the mass step is conventionally placed at $10^{10}\,\Msun$~\citep{Sullivan2010}, preferred split locations vary across $10^{9.73}$--$10^{10.8}\,\Msun$~\citep{Kelly2010,Kelsey2021,Ponder2021,Thorp2022} and are treated as a cosmological systematic~\citep{Brout2022,Vincenzi2024,Popovic2026DES}. Scanning from the 5th to 95th percentiles of host mass ($10^{8.2}$--$10^{10.9}\,\Msun$) and $g-z$ colour shows that $(\Mej, A_V, (B-V)_0)$ remains smaller than SALT3 at every mass split (by $0.014$--$0.054$\,mag) and for all colour splits bluer than $g-z = 1.3$, while the four-parameter set tracks SALT3. Against local host properties (718 events), $(\Mej, A_V, (B-V)_0)$ similarly lowers the local mass step above $10^{10}\,\Msun$ from $-0.113\pm0.022$ to $-0.070\pm0.021$\,mag (paired reduction $0.043$\,mag, $95\%$ interval $0.018$--$0.069$; at the median local mass of $10^{8.90}\,\Msun$, from $-0.067$ to $-0.050\pm0.013$\,mag) and the median local $g-z$ step ($1.044$\,mag) from $-0.086$ to $-0.070\pm0.013$\,mag (paired reduction $0.015$\,mag), while the four-parameter model again matches SALT3 ($-0.087\pm0.013$\,mag).

To see which parameters drive the host-mass step, we decompose each standardisation into the unstandardised step plus the contribution of each predictor. For SALT3, the unstandardised step ($+0.046$\,mag) becomes $-0.084$\,mag through stretch $x_1$ ($-0.115$\,mag) and colour $c$ ($-0.014$\,mag). In $(\Mej, A_V, (B-V)_0)$, the unstandardised step ($+0.060$\,mag) becomes $-0.069$\,mag through $\Mej$ ($-0.029$\,mag) and intrinsic $(B-V)_0$ ($-0.100$\,mag), with $A_V$ contributing $0.000$\,mag under our common host $R_V = 1.85$. In natural units for $(t_{\rm d}, A_V, (B-V)_0)$, the slopes cleanly separate dust ($+1.37$ per unit $A_V$, corresponding to $\beta_{\rm dust} \simeq 2.52$ at $R_V = 1.85$), intrinsic colour ($\beta_{\rm int} = +2.65$, rising to $+3.12$ with slope $-1.54$ on $(r-i)_0$), and width ($-0.49$\,mag per unit $\ln t_{\rm d}$, or $-0.18$\,mag per unit $\ln\Mej$).

\subsection{Distance information in the standardisation parameters}
\label{sec:leak}

To verify that distance information does not leak into our standardisation predictors, we ran two tests. First, shifting the observed fluxes of 150 ZTF SNe by random grey offsets ($0.2$\,mag dispersion) and refitting at their redshift distances shows that model peak $m_B$ tracks the offset cleanly (slope $0.986$, scatter $0.006$\,mag), whereas $\Mni$ absorbs only $86\%$ because fixed-distance fits push part of a grey offset into $A_V$ ($+0.11$ per unit magnitude offset). Standardising $\Mni$ directly would therefore borrow distance information through dust (artificially lowering scatter from $0.187$ to $0.148$\,mag), and even with $m_B$ as the target, distance-fixed predictors still absorb $8$--$20\%$ of an injected offset. In our redshift-blind run ($\sigma_{\mu,i} = 3$\,mag), however, only $2.5\%$ ($2.0$--$3.1\%$) of a brightness shift reaches $\Mni$, and within each blind posterior the standardisation prediction shifts by $0.07$ ($0.05$--$0.09$) per unit $\ln\Mni$. To first order, distance information therefore leaks at only $\simeq 0.2\%$ ($0.002$ per unit magnitude offset, biasing a $0.1$\,mag step by at most $0.0002$\,mag), though we note this bound uses posterior covariances rather than a full blind-chain rerun.

\subsection{Properties of the learned SED}
\label{sec:sed_learned}

\begin{figure*}
\centering
\includegraphics[width=\textwidth]{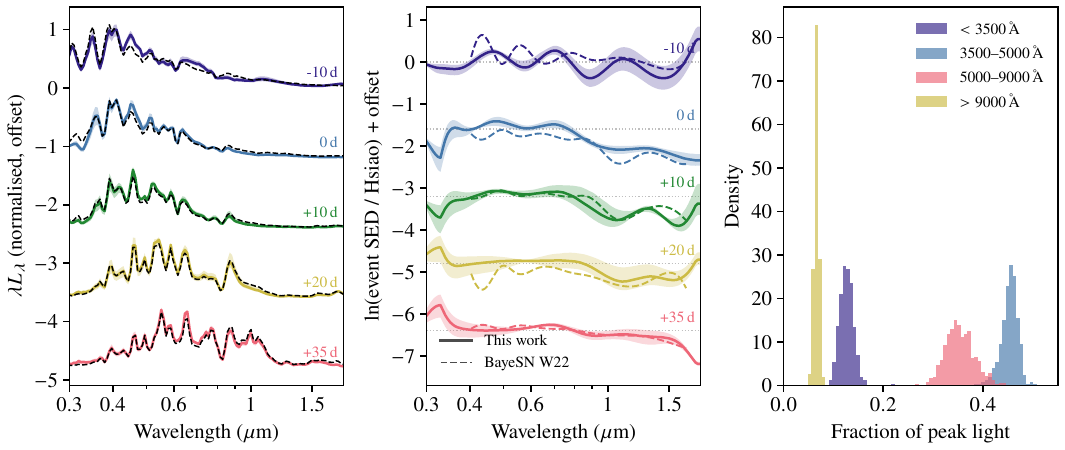}
\caption{Learned intrinsic SED of the 845 fitted ZTF SNe. Left: median over events of each SN's own SED (solid curves with 16th--84th percentile bands) at five rest-frame phases compared against the \citet{Hsiao2007} template (dashed), vertically offset by phase and normalised over $0.33$--$1.8\,\mu$m for display (the forward model normalises over the full $100$--$300\,000$\,\AA\ domain, equation~\ref{eq:normalisation}). Centre: log-ratio of the learned event SEDs to Hsiao (solid, with 16th--84th percentile bands) alongside the BayeSN W22 mean intrinsic SED at $\theta=0$ relative to Hsiao (dashed, shown over $0.4$--$1.7\,\mu$m where constrained by its training data), offset vertically by phase around dotted zero-difference lines; both models independently recover the same near-infrared deficit relative to Hsiao. Right: distributions across the 845 ZTF SNe of the fraction of each event's peak bolometric light falling in four wavelength intervals ($<3500$\,\AA, $3500$--$5000$\,\AA, $5000$--$9000$\,\AA, and $>9000$\,\AA; the four fractions sum to unity for each SN).}
\label{fig:sed}
\end{figure*}

Figure~\ref{fig:sed} shows the learned intrinsic SED across the 845 fitted ZTF SNe. Across the optical ($0.4$--$0.9\,\mu$m), the median event SED stays within $0.4$ in $\ln L_\lambda$ of the \citet{Hsiao2007} template from $-10$ to $+35$\,d. The main departures are in the near-ultraviolet (fainter than Hsiao by up to $0.46$ in $\ln L_\lambda$ near peak, brighter past $+20$\,d) and beyond $1\,\mu$m (up to a factor of two fainter from maximum light onwards).

This near-infrared deficit is not unique to our model: BayeSN~\citep[model W22;][]{Mandel2022,Ward2023}, which likewise warps the \citet{Hsiao2007} template on optical and near-infrared photometry, recovers the same trend. Relative to Hsiao at peak, our mean $V-J$ and $V-H$ colours are bluer by $0.59$ and $0.81$\,mag (vs.\ $0.57$ and $0.73$\,mag for BayeSN W22), and both models agree in $V-J$ and $V-H$ within $0.14$\,mag through $+35$\,d (differing by $0.07$--$0.31$ in $\ln L_\lambda$ rms across $0.4$--$0.9\,\mu$m; below $0.35\,\mu$m BayeSN's mean is unconstrained). Across ZTF at peak, individual SEDs vary by $0.17$ in $\ln L_\lambda$ (median $68\%$ width) and allocate $13\%$ of bolometric flux below $3500$\,\AA\ ($6.4\%$ below $3000$\,\AA, within the observed $3$--$9\%$ for normal SNe~Ia and avoiding blackbody UV over-extrapolation), $46\%$ over $3500$--$5000$\,\AA, $35\%$ over $5000$--$9000$\,\AA, and $7\%$ above $9000$\,\AA; optical-only ZTF photometry leaves the near-infrared component weight at its prior ($0.0101\pm0.0001$).

\subsection{Uncertainty from SED training}
\label{sec:sed_uncertainty}

As the SED is frozen after Stage 1, we test sensitivity to the 46-event CSP training set by continuing SED training (steps 4500--6000) on four bootstrap resamples and the original run, then refitting a 121-event random ZTF subsample under each SED. The physical host-mass relation is unchanged: the $\Mej$ offset at fixed $x_1$ stays within $-0.07$ to $-0.02\,\Msun$ ($\pm0.07\,\Msun$, vs.\ $-0.04\pm0.06\,\Msun$ baseline). Individual SED scores, however, shift by up to $2\sigma$ even on the unperturbed sample, confirming they act as a flexible basis rather than unique physical observables. Appending 166 additional low-redshift SNe~Ia from CfA, Foundation, and nearby optical/NIR compilations~\citep{Ward2023,Grayling2026b} yields nearly identical ZTF standardisation on the same 121 events ($0.160$\,mag wRMS, $\sigma_{\rm int} = 0.144$\,mag, mass step $-0.088\pm0.031$\,mag, within $1\sigma$ of the four CSP bootstraps at $0.166\pm0.005$, $0.150\pm0.006$, and $-0.084\pm0.016$\,mag), showing that expanding the training set does not tighten optical-only ZTF standardisation.

We also tested varying the number of global spectral modes $K$ in equation~(\ref{eq:sed}). By singular-value decomposition across $-10$ to $+35$\,d ($0.33$--$1.8\,\mu$m), two orthogonal modes carry $96.9\%$ of the squared mode variance ($80.6\%$ and $16.3\%$; mode 3 carries $3.1\%$). Retraining CSP (steps 4500--6000, fixed physics) with $K = 0, 1, 2$, and $3$ gives final photometric log-likelihoods across 7071 $u$--$H$ observations of $-71\,452$, $+2\,074$, $+12\,900$, and $+19\,111$: $K \le 1$ fails badly, and $K = 3$ improves on $K = 2$ by $\Delta\ln\mathcal{L} = +6\,211$ ($\simeq 0.88$ per observation, far exceeding the $\sim 950$ BIC penalty). Applied redshift-blind to the 121 ZTF events, all three ($K = 1, 2, 3$) recover the same host-mass step ($-0.086\pm0.032$, $-0.087\pm0.030$, and $-0.084\pm0.034$\,mag for $(\Mej, A_V, (B-V)_0)$, vs.\ $-0.074\pm0.032$\,mag for SALT3), while on optical-only $gri$ photometry $K = 2$ matches or slightly betters $K = 3$ in scatter ($0.158$\,mag wRMS and $0.128$\,mag NMAD vs.\ $0.177$ and $0.149$\,mag for retrained $K = 3$, and $0.166$ / $0.157$\,mag for SALT3; $0.153$ vs.\ $0.177$\,mag for the four-parameter set). Thus two modes suffice for optical-only standardisation, whereas nine-band CSP photometry resolves the third optical--NIR mode.

\section{Discussion}
\label{sec:discussion}

\subsection{Relation to empirical models, dust, and host environment}
\label{sec:empirical}

Empirical models such as SALT3 and hierarchical SED models such as BayeSN achieve low Hubble scatter without explicit explosion physics, whereas simple blackbody-SED physical models~\citep{Sarin2026b} over-extrapolate flux below $3000$\,\AA, inflating $\Lbol$, $\Mni$, and $f_{\rm Ni}$. Replacing a blackbody SED with an energy-conserving CSP-trained surface ($6.4\%$ of peak light below $3000$\,\AA) lowers the ZTF median $\Mni$ from $\approx 1.2$ to $0.57\,\Msun$, restores realistic peak luminosities and $t_0$ scales~\citep{Sharon2020,Sharon2020b,Sharon2025}, and cuts the host-mass $\Mni$ offset by more than a factor of three, consistent with the simplified SED being a major contributor to the $^{56}$Ni bias (albeit with large uncertainties which are well controlled by hierarchical modelling~\citep{Sarin2026b}).

In our model, the lower ejecta masses we infer in massive hosts track the well-known host dependence of SALT3 stretch~\citep{Howell2009,Nicolas2021,Ginolin2025a}: high- and low-mass host SNe follow the same $\Mej$--$x_1$ relation, with no remaining ejecta-mass offset once $x_1$ is fixed. Conversely, at fixed $\Mej$, SNe in massive hosts still have lower $x_1$ because empirical optical stretch reflects intrinsic colour and cooling in addition to ejecta mass.
Standardising on $(\Mej, A_V, (B-V)_0)$ matches SALT3 scatter while lowering the mass step by $0.020$\,mag to $-0.073\pm0.013$\,mag, and $(t_{\rm d}, A_V, (B-V)_0, (r-i)_0)$ lowers scatter below SALT3 ($0.173$\,mag wRMS, $0.152$\,mag NMAD) with a $-0.088\pm0.013$\,mag step. As we use a single host dust law ($R_V\simeq 1.85$), extinction contributes $0.000$\,mag to the step. One explanation is that the step has both intrinsic and extrinsic components; some previous analyses find that environment-dependent dust laws contribute to the mass step~\citep[e.g.][]{Brout2021,Popovic2023,Grayling2025}. However, a dust-driven step requires large $R_V$ differences between high- and low-mass galaxies, and a recent BayeSN analysis of this ZTF sample still finds an intrinsic host-mass step of $0.103\pm0.018$\,mag with no significant $R_V$ difference~\citep{Ginolin2026} (samples and cuts differ, so we do not compare amplitudes directly). Taken together, modelling these physical differences under a single dust law does not eliminate the step.

A striking feature of Table~\ref{tab:tripp} and Fig.~\ref{fig:step_sizes} is that standardising on physical timescales and dust alone, $(t_{\rm d}, A_V)$ or $(t_{\rm d}, t_\gamma, A_V)$, leaves the host-mass step consistent with zero ($+0.002\pm0.021$ and $+0.007\pm0.020$\,mag, flat across host mass), whereas the $-0.07$ to $-0.09$\,mag step appears only once peak intrinsic colour $(B-V)_0$ is added to tighten the scatter from $0.277$ to $0.175$\,mag. At fixed $t_{\rm d}$, SNe~Ia in massive hosts are intrinsically redder by $\Delta(B-V)_0 = +0.032\pm0.006$\,mag ($5.3\sigma$), persisting at $+0.045\pm0.010$\,mag ($4.4\sigma$) even for $A_V < 0.15$\,mag, ruling out a single-$R_V$ dust artefact (SALT3 $x_1$ absorbs this shift at $r = -0.69$ at fixed $t_{\rm d}$, hiding the separation). Two interpretations fit our optical-only ZTF data: either high- and low-mass host SNe share the same mean peak luminosity at fixed $(t_{\rm d}, A_V)$ and a luminosity-neutral spectral shift (such as metallicity-driven iron-group blanketing) reddens $(B-V)_0$ so that applying the global $\beta_{\rm int} = 2.65$ slope induces a negative mass step; or omitting $(B-V)_0$ leaves $0.256$\,mag of $^{56}$Ni scatter uncorrected, allowing lower $^{56}$Ni in massive hosts to cancel an independent negative step near zero in mass (consistent with the $+0.034$ to $+0.040\pm0.020$\,mag host-colour step remaining for $(t_{\rm d}, A_V)$). Conditioning near-infrared or direct bolometric luminosities on $t_{\rm d}$ without optical colour or stretch, together with peak ultraviolet/blue spectroscopy and larger samples, will separate these two possibilities.

The step may also reflect ejecta kinematics and explosion channels: \citet{Burgaz2026} find on ZTF DR2 that the step drops from $0.149\pm0.024$\,mag for normal-velocity SNe~Ia to $0.046\pm0.041$\,mag above $v_{\rm Si} = 12\,000\,{\rm km\,s^{-1}}$~\citep[see also][]{Foley2011,Foley2011_vel,Siebert2020}, while spectral fits by \citet{Magee2026} find that roughly two-thirds of ZTF DR2 SNe~Ia favour sub-Chandrasekhar-mass explosions and that Chandrasekhar-mass explosions show more negative Hubble residuals. Our framework allows such velocity and channel effects to enter the forward physics directly via equation~(\ref{eq:velocity}).

\subsection{$^{56}$Ni and distance}
\label{sec:nickel_distance}

As \snmix{}'s grey optical opacity depends only on time rather than temperature, the bolometric light curve scales strictly linearly with $\Mni$ at fixed ejecta structure~\citep[unlike real SNe~Ia, where radioactive heating controls ionisation and post-peak decline;][]{Pinto2000,Kasen2006,Wygoda2019}. Consequently, $\Mni$ is strictly degenerate with luminosity distance and cannot serve as a distance-free standardisation parameter: as Sec.~\ref{sec:leak} shows, standardising redshift-distance $\Mni$ borrows brightness information through $A_V$, whereas standardising peak $m_B$ from the redshift-blind run avoids this bias. A stable, differentiable temperature-dependent opacity is needed to give $\Mni$ a direct light-curve shape signature. Our StaNdaRT tests showed unstable transition temperatures across folds without changing $\Mni$, so we leave this to future work.

\subsection{Limitations and generality}
\label{sec:limitations}

Several caveats apply when interpreting absolute parameter values:
\begin{itemize}
\item \emph{Phase window and peak overshoot:} past $+40$\,d, peak-fitted \snmix{} light curves overshoot CSP quasi-bolometric curves by $10$--$14\%$ (extending to $+60$\,d shifts $\Mej$ by $\simeq 25\%$), and against StaNdaRT \snmix{} generally peaks brighter. 
Both introduce shared systematics on absolute $\Mej$ (consistent with $t_0$ sitting a few days above \citealt{Sharon2025}), while differential host comparisons at fixed phase window remain insulated. 
Incorporating late-time Katz-integral $t_0$ measurements~\citep{Katz2013,Sharon2020,Sharon2020b,Sharon2025} would complement Si\,\textsc{ii} velocities where late bolometric data exist.
\item \emph{Uncertainty calibration and dust:} pre-velocity synthetic ejecta-mass intervals were $20$--$30\%$ narrower than nominal, implied CSP bulk speeds sit $\simeq 10\%$ above Si\,\textsc{ii} line velocities, and our hierarchy adopts a single common host $R_V$ and fixed Stage-1 point-estimate SED rather than propagating full uncertainty on the SED and population parameters.
\item \emph{Sample scope:} CSP training requires $s_{BV}\ge 0.8$ and ZTF covers spectroscopically normal SNe~Ia at $z<0.06$; extending to fast decliners and $z>0.06$ remains for future work.
\end{itemize}

More broadly, separating a physical bolometric engine from an energy-normalised learned SED does not depend on the Hsiao template or normal-SN~Ia priors: for other transient classes, the same structure can pair competing physical power sources with alternative spectral priors and joint photometric--spectroscopic constraints.

\section{Conclusions}
\label{sec:conclusions}

We have introduced a differentiable physical population model for SNe~Ia combining multi-shell radiation diffusion (\snmix{}) with a learned, energy-normalised SED, trained on CSP optical and near-infrared photometry and applied to 845 normal ZTF DR2 SNe~Ia at $z<0.06$. Our main conclusions are:
\begin{enumerate}
\item Phase-by-phase energy normalisation ties every model photon to radioactive decay. Against StaNdaRT radiative-transfer benchmarks, \snmix{} recovers $\Mni$ within $2$--$6\%$ across $3$--$60$\,d (well inside the $11$--$27\%$ inter-code spread).
\item Si\,\textsc{ii} velocities break the broadband ejecta-mass--velocity degeneracy, cutting CSP events above $2.1\,\Msun$ from 35 to 2 at $<0.01$ in $\chi^2$ per observation. Across ZTF, we infer hierarchical population medians of $\Mej = 1.39\,(1.02\text{--}1.72)\,\Msun$, $\Mni = 0.57\,(0.43\text{--}0.71)\,\Msun$, $\Mni/\Mej = 0.42$, and $f_{\rm mix} = 0.59$.
\item SNe~Ia in massive hosts ($\log_{10}(M_*/\Msun)>10$) have lighter ejecta by $0.20\pm0.03\,\Msun$, slightly less $^{56}$Ni ($0.04\pm0.01\,\Msun$), and indistinguishable $A_V$ under a common dust law. The $\Mej$ difference vanishes at fixed SALT3 $x_1$ ($+0.04\pm0.03\,\Msun$), while $x_1$ retains an additional host-mass correlation at fixed $\Mej$ linked to intrinsic colour.
\item Redshift-blind physical standardisation with $(\Mej, A_V, (B-V)_0)$ matches SALT3 scatter on 727 ZTF SNe ($0.177$ vs.\ $0.180$\,mag wRMS; $0.161$ vs.\ $0.160$\,mag NMAD) while lowering the host-mass step from $-0.093\pm0.014$ to $-0.073\pm0.013$\,mag (holding across all split thresholds and local host tracers); $(t_{\rm d}, A_V, (B-V)_0, (r-i)_0)$ pushes scatter below SALT3 ($0.173$\,mag wRMS, $0.152$\,mag NMAD) with a $-0.088\pm0.013$\,mag step. Standardising on $(t_{\rm d}, A_V)$ alone leaves the host-mass step near zero ($+0.002\pm0.021$\,mag at $0.277$\,mag wRMS), with the full step emerging once $(B-V)_0$ is included because massive-host SNe are intrinsically redder at fixed $t_{\rm d}$ ($\Delta(B-V)_0 = +0.032\pm0.006$\,mag, persisting at $A_V < 0.15$\,mag).
\end{enumerate}

As optical-only ZTF photometry is limited by the light curves themselves rather than SED training size (quadrupling the training set leaves standardisation unchanged), breaking the remaining intrinsic-colour and dust degeneracies will require broader wavelength coverage of target SNe. Upcoming optical-to-near-infrared surveys with Rubin~\citep{Ivezic2019} and Roman~\citep{Hounsell2018,Rose2021} will enable differentiable physical population models like this one to test directly how explosion physics, spectral evolution, and host environments evolve with redshift.

\section*{Acknowledgments}

We thank Doron Kushnir for helpful discussions on ultraviolet SED extrapolation and gamma-ray escape timescales. NS is supported by the Kavli Foundation. MG is supported by the European Union's Horizon 2020 research and innovation programme under ERC Grant Agreement No.~101002652 (BayeSN; PI K.~Mandel). LK acknowledges support for an Early Career Fellowship from the Leverhulme Trust through grant ECF-2024-054 and the Isaac Newton Trust through grant 24.08(w). This work made use of the OzSTAR national facility at Swinburne University of Technology. The OzSTAR program receives funding in part from the Astronomy National Collaborative Research Infrastructure Strategy (NCRIS) allocation provided by the Australian Government. The authors acknowledge the use of Claude Opus 5.5 (Anthropic) and Gemini 4 Neon (Google) to assist with analysis code, plotting scripts and drafting of the text. The authors take full responsibility for the content. This work made use of \program{JAX}~\citep{Bradbury2018}, \program{NumPy}~\citep{Harris2020}, and \program{Matplotlib}~\citep{Hunter2007}.

\section*{Data Availability}

The CSP and ZTF photometric datasets used in this work are publicly available~\citep{Krisciunas2017,Rigault2025a}. The per-event posterior summaries and plots are available at \url{https://nikhil-sarin.github.io/snia-physical-sed-catalogue/}. Full posteriors are available upon reasonable request to the corresponding author. 

\bibliographystyle{mnras}
\bibliography{refs}

\bsp
\label{lastpage}
\end{document}